\documentclass[sort&compress,number,5p]{elsarticle}
\usepackage{amsfonts}
\usepackage{amssymb}
\usepackage{amsmath}
\usepackage{hyperref}
\usepackage{graphicx}
\usepackage{xcolor}
\usepackage{soul}
\begin{document}
\begin{frontmatter}
\title{\mbox{Super-Hydrophobic Multi-Walled Carbon Nanotube Coatings for Stainless Steel}}
\author[tv,infn]{Francesco De Nicola\corref{cor1}}
\ead{fdenicola@roma2.infn.it}
\ead[url]{0039 0672594532}
\author[tv,infn]{Paola Castrucci}
\author[tv,infn]{Manuela Scarselli}
\author[ing]{Francesca Nanni}
\author[instm]{Ilaria Cacciotti}
\author[tv,infn,ism]{\mbox{Maurizio De Crescenzi}}
\cortext[cor1]{Corresponding author}
\address[tv]{Dipartimento di Fisica, Universit\'a di Roma Tor Vergata, Via della Ricerca Scientifica 1, 00133 Roma, Italy}
\address[infn]{Istituto Nazionale di Fisica Nucleare, Universit\'a di Roma Tor Vergata (INFN-Roma Tor Vergata), Via della Ricerca Scientifica 1, 00133 Roma, Italy}
\address[ing]{Dipartimento di Ingegneria dell'Impresa, Universit\'a di Roma Tor Vergata (INSTM-UdR Roma Tor Vergata), Via del Politecnico 1, 00133 Roma, Italy}
\address[instm]{Universit\'a di Roma Niccol\'o Cusano (INSTM-UdR), Via Don Carlo Gnocchi 3, 00166 Roma, Italy}
\address[ism]{Istituto di Struttura della Materia, Consiglio Nazionale delle Ricerche (ISM-CNR), Via del Fosso del Cavaliere 100, 00100 Roma, Italy}
\begin{abstract}
We have taken advantage of the native surface roughness and the iron content of AISI 316 stainless steel to direct grow multi-walled carbon nanotube (MWCNT) random networks by chemical vapor deposition (CVD) at low-temperature ($< 1000^{\circ}$C), without the addition of any external catalysts or time-consuming pre-treatments. In this way, super-hydrophobic MWCNT films on stainless steel sheets were obtained, exhibiting high contact angle values ($154^{\circ}$) and high adhesion force (high contact angle hysteresis). Furthermore, the investigation of MWCNT films at scanning electron microscopy (SEM) reveals a two-fold hierarchical morphology of the MWCNT random networks made of hydrophilic carbonaceous nanostructures on the tip of hydrophobic MWCNTs. Owing to the Salvinia effect, the hydrophobic and hydrophilic composite surface of the MWCNT films supplies a stationary super-hydrophobic coating for conductive stainless steel. This biomimetical inspired surface not only may prevent corrosion and fouling but also could provide low-friction and drag-reduction.
\end{abstract}
\end{frontmatter}
\section{Introduction}
\label{sec:Introduction}
\indent Super-hydrophobic surfaces (i.e., water contact angle greater than 150$^{\circ}$) have attracted recently much attention in fundamental research \cite{Li2002,Sun2005,Shibuichi1996,Huang2012,Furstner2005} and potential industrial applications, such as waterproof surfaces \cite{Lau2003}, anti-sticking \cite{Wang2007}, anti-contamination \cite{XingJiuHuang2007}, self-cleaning \cite{Furstner2005}, anti-fouling \cite{Zhang2005}, anti-fogging \cite{Lai2012}, low-friction coatings \cite{Jung2009}, adsorption \cite{Li2010a}, lubrication \cite{Adamson1997}, dispersion \cite{Gennes2003}, and self-assembly \cite{Huang2012}. In general, artificial super-hydrophobic surfaces can be realized governing both the chemical compositions and morphological structure of the solid surfaces. In particular, surface roughness \cite{Wenzel1936,Koch2009} (micro- and nano-morphology) may also be enhanced especially by hierarchical \cite{Feng2008,Sun2005,XingJiuHuang2007,Jung2009,Egatz-Gomez2012,Bittoun2012,Koch2009,Barthlott2010} and fractal structures \cite{Shibuichi1996,Bittoun2012}, possibly allowing air pocket formation to further repel water penetration \cite{Cassie1944}. Nevertheless, realizing a permanent super-hydrophobic surface remains quite a challenge. Lately, chemical \cite{Wang2011}, mechanical \cite{Jung2009}, thermal stability \cite{Wang2008}, and time durability \cite{Huang2005} have been addressed.\\
\indent However, the best and most efficient surfaces known so far evolved in 460 million years in plants and animals owe to adaptation to different environments and now they serve as models for the development of artificial biologically inspired or biomimetic materials \cite{Bhush2009,Bar-Cohen2005,Koch2009}. Recent studies demonstrate that super-hydrophobicity of many natural surfaces \cite{Feng2008,Sun2005,Koch2009,Barthlott2010} principally results from the presence of at least two-fold morphology at both micro- and nano-scales and the low energy materials on the surfaces. For instance, the hierarchical architecture of the \textsl{Salvinia} leaf surface is dominated by complex elastic papillae millimetric in size coated with self-assembly nano-scaled epidermal wax crystals ranging in sizes from 0.2 to 100 $\mu$m \cite{Barthlott2010,Koch2009}. The terminal cells of each super-hydrophobic papilla lack the wax crystals and form evenly distributed hydrophilic cells that cover only 2\% of the surface \cite{Barthlott2010}. These hydrophilic cells stabilize the air layer by pinning the liquid-vapor interface to the tips of the papillae. This prevents the loss of air caused by formation and detachment of gas bubbles due to instabilities, such as pressure fluctuations, especially in a turbulent water flow environment. The unique combination of hydrophilic cells on super-hydrophobic papillae provide a promising concept for the development of a coating with a long-term super-hydrophobic behaviour. In general, the adhesion with the water is so strong that the elastic papillae bend and swing back when the tips snap off the droplets \cite{Barthlott2010}.\\
\indent The so-called Salvinia or petal effect \cite{Feng2008,Koch2009,Barthlott2010} is therefore referred to super-hydrophobic adhesive surfaces with hydrophilic and hydrophobic hierarchical morphology providing sufficient roughness for exhibiting both a large contact angle and a high contact angle hysteresis, conversely to the lotus effect \cite{Sun2005,Feng2008} (high contact angle value and low contact angle hysteresis). Consequently, a water droplet on such a surface is nearly spherical in shape and cannot roll off even when the leaf is turned upside down. However, larger drops can roll off the surface at the slightest tilting or vibration \cite{Decker1996}.\\
\indent In general, it is very difficult to fabricate an applicable engineering super-hydrophobic surface on stainless steel, because the textured films easily fall off from the stainless steel substrate. Lately, some achievements on the realization and characterization of stable super-hydrophobic surfaces on stainless steel have been made \cite{Chen2008,Yang2010b,Shen2005,Satyaprasad2007} and particularly using carbon nanotube coatings \cite{Lee2010,Lee2011}. Furthermore, stainless steel potential applications include electrodes for super-capacitors \cite{Chen2011}, fuel cells \cite{Chen2012}, capacitive deionization \cite{Anderson2010} and capacitive mixing for extracting energy from salinity difference of water resources \cite{Brogioli2009}, field emission probes \cite{Wang2001}, sensors \cite{Minnikanti2009}, catalyst support for wastewater treatment \cite{Sano2012} and tribological applications \cite{Abad2008}. Therefore, stainless steel may be considered as a valid candidate for direct growth \cite{Hashempour2013} of carbon nanotubes by CVD, also because of its high content of iron as the catalyst element. In particular, direct growth is widely used due to several advantages, such as capability to produce dense and uniform deposits, reproducibility, strong adhesion, adjustable deposition rates, ability to control crystal structure, surface morphology and orientation of the CVD products, reasonable cost and wide scope in selection of chemical precursors.\\
\indent Recently, we have shown \cite{Camilli2011,Camilli2012,Camilli2012a,Camilli2013} that the direct growth of high quality MWCNTs on stainless steel in the absence of any external catalysts is possible. Moreover, acid treatments and oxidation-reduction stages on this type of surface are not necessary because of the native nano-scale roughness of the substrate and the iron-rich substrate surface both act as an efficient catalyst or template in the synthesis of MWCNTs. Particularly, at  our working temperature mostly iron nanoparticles are involved in the growth mechanism. Furthermore, after the first growth, the stainless steel substrate may be used again, just carefully removing the synthesized carbon nanotubes in an ultrasonic bath. We remark that ultrasonication is generally needed to detach the MWCNT film from the steel substrate, due to its strong adhesion \cite{Camilli2011,Hashempour2013}.\\
\indent Here, we illustrate a simplified recipe to synthesize MWCNTs on a sheet of AISI 316 stainless steel by CVD, without any external catalysts. Moreover, we will investigate the MWCNT hierarchical morphology from the SEM micrographs of the films and their super-hydrophobic properties will be characterized. In particular, we will show that owing to their particular hierarchical architecture, the super-hydrophobic MWCNT coatings for stainless steel exhibit long-term high contact values and also high adhesive force with water (high contact angle hysteresis). Therefore, the super-hydrophobic state achieved is stationary.
\section{Experimental}
\label{sec:exp}
\indent A $30\times40$ mm$^{2}$ piece of AISI 316 stainless steel sheet (Fe 70\%, Cr 18\%, Ni 10\%, and Mo 2\%, Goodfellow Cambridge, Ltd.) was carefully sonicated in deionized water and degreased in isopropyl alcohol for 10 min. Then, the steel substrate was placed on a molybdenum sample holder acting also as resistive heater and inserted into an ultra high vacuum chamber and the pressure was brought up to $10^{-2}$ Torr by a rotary pump. At this stage, argon gas (500 sccm) was inserted at 12 Torr and then the heater was increased at the working temperature $\approx730$ $^{\circ}$C. The sample temperature was controlled with an optical pyrometer, so when the substrate reached the working temperature, acetylene (C$_{2}$H$_{2}$) was introduced (200 sccm) in the chamber and MWCNTs grew in dynamic condition, since the rotary pump was kept in operation during the process. After 10 min of growth, Ar gas (500 sccm) was inserted in chamber for 5 min.
\section{Results and discussion}
\label{sec:results}
\begin{figure*}[ht]
	\centering
		\includegraphics[width=1.00\textwidth]{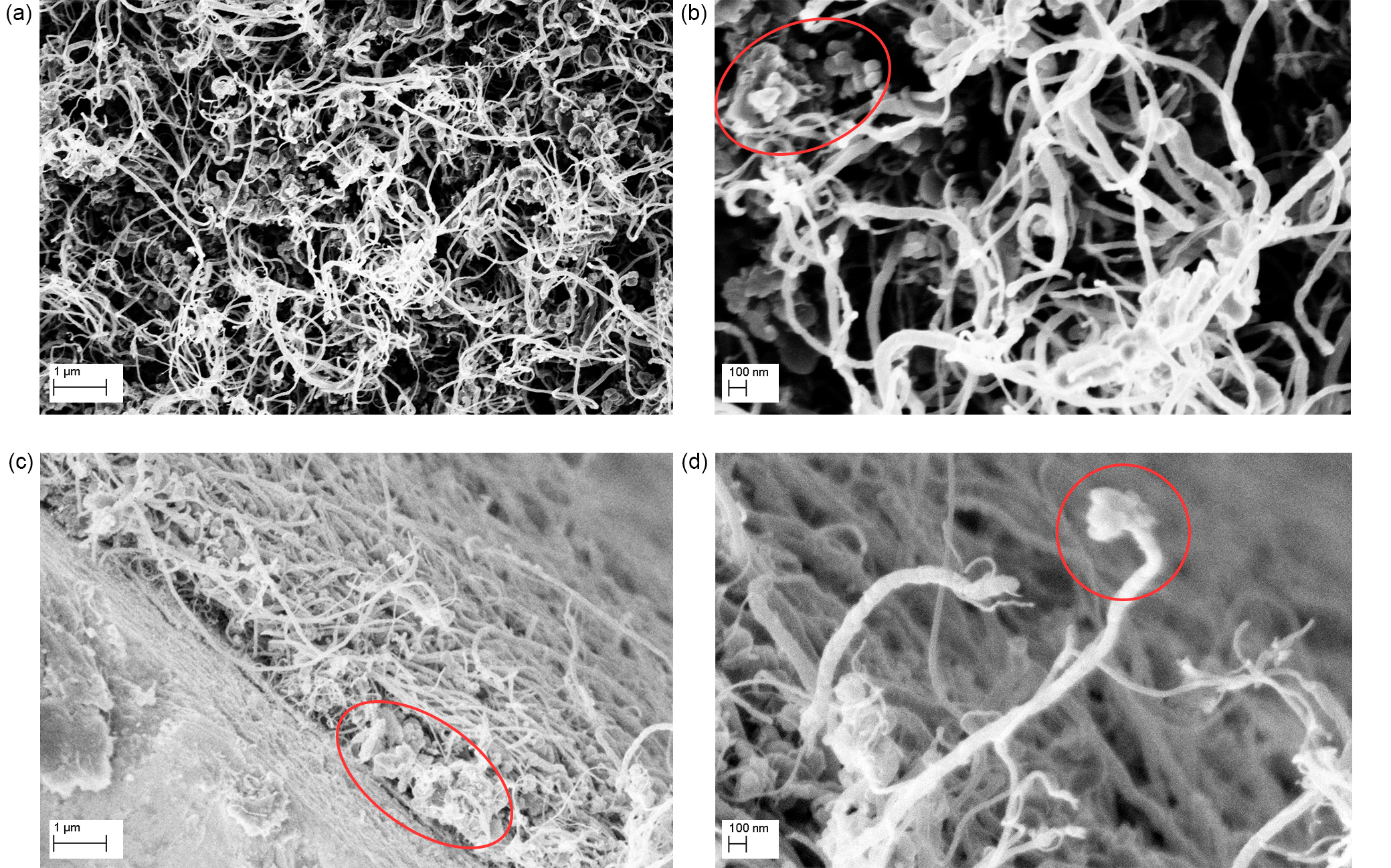}
	\caption{Scanning electron microscopy images of MWCNTs direct grown by CVD on stainless steel without any external catalysts or pre-treatments, at different magnifications, 30,000$\times$ (a,c) and 100,000$\times$ (b,d). Carbonaceous nanostructures may be seen close to the stainless steel surface and around MWCNT tips (red circles).}
	\label{fig:Figure1}
\end{figure*}
\subsection{Microscopic characterization of MWCNT films}
\label{sec:SampleCharacterization}
\indent In Figure \ref{fig:Figure1}a,d SEM (Zeiss Leo Supra 35) images of our synthesized nanostructures are reported, showing that a high density of randomly oriented MWCNTs uniformly grew on the stainless steel sheet with an average film thickness of $1.6\pm0.8$ $\mu$m. Moreover, MWCNTs come with a wide distribution of tube diameters, with average value $58.19\pm18.35$ nm. Also, we have shown in our past works \cite{Camilli2011,Camilli2012,Camilli2012a,Camilli2013} transmission electron microscopy (TEM) images confirming the multi-walled nature of the as-grown carbon nanotubes. Furthermore, in Figure \ref{fig:Figure1}b,d it may be observed that MWCNTs are mostly capped and often they present carbonaceous nanostructures (amorphous and/or graphitic carbon) around the tips (Figure \ref{fig:Figure1}d) and close to the stainless steel surface (Figure \ref{fig:Figure1}c), with a characteristic dimension of hundred nanometers, as also reported by other authors \cite{Hashempour2013}. 
\begin{figure*}[ht]
	\centering
		\includegraphics[width=1.00\textwidth]{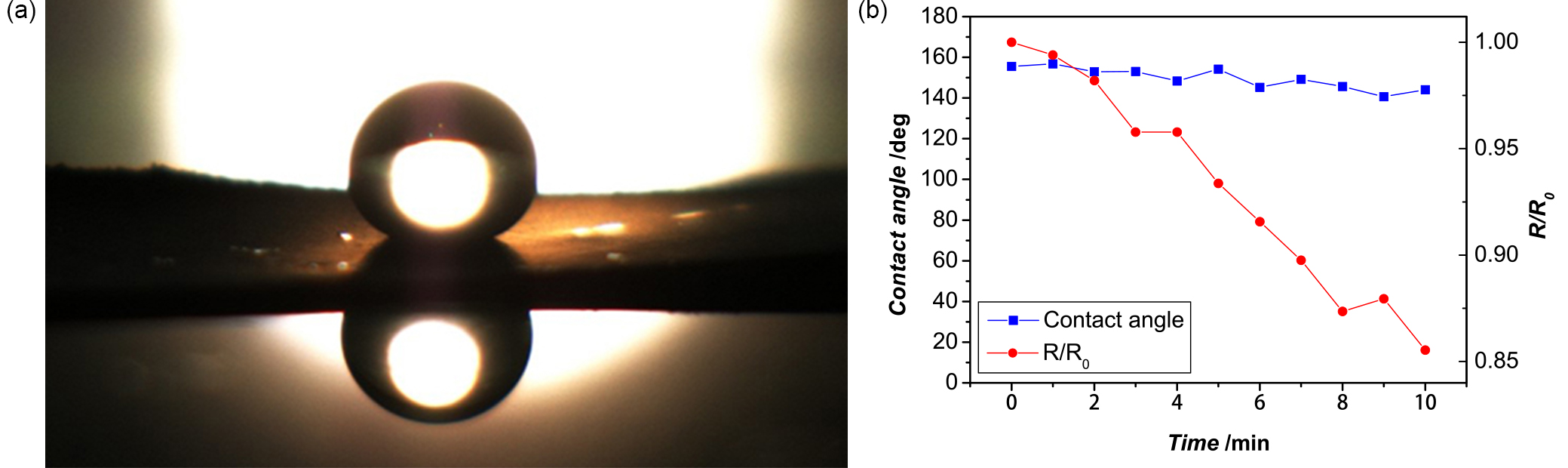}
	\caption{(a) The water contact angle value measured on the MWCNT films deposited on stainless steel is $\Theta=154^{\circ}\pm4^{\circ}$. (b) Variations of contact angle and drop radius as functions of the elapsed time from drop cast on the porous MWCNT films. The constant trend of the contact angle value proves the stability in time of the super-hydrophobic state, while the linear decrease of droplet radius is due to the liquid evaporation and not to the suction process.}
	\label{fig:Figure2}
\end{figure*}
\subsection{Wetting properties of MWCNT films}
\label{sec:ContactAngleMeasurements}
\indent Moreover, we characterized the wettability of MWCNT films acquiring images of sessile water drops cast on the carbon nanotube films by a custom setup with a CCD camera. Static advanced contact angles were measured increasing the volume of the drop by step of 1 $\mu$L. A plugin \cite{Stalder2010} for the open-source software ImageJ was exploited to estimate the contact angle values by using cubic B-Spline interpolation of the drop contour to reach subpixel resolution, with an accuracy of $0.02^{\circ}$. The deionized water (18.2 M$\Omega$\:cm) drop volume used to achieve the contact angles of samples was 10 $\mu$L. Moreover, every contact angle was measured $15$ s after drop casting to ensure that the droplet reached its equilibrium position. In Figure \ref{fig:Figure2}a the image of a water droplet cast on the MWCNT film is shown. The experimental contact angle value is $\Theta=154^{\circ}\pm4^{\circ}$ with no observable roll-off angle, even if the substrate is turned upside down. Therefore, we infer that the contact angle hysteresis is so high to pin the water droplet on the MWCNT surface. The adhesive force in unit of length of a surface in contact with water is given by \cite{Adamson1997}
\begin{equation}
	F_{adh}=\gamma_{LV}\left(1+\cos{\Theta}\right),
\end{equation}
where $\gamma_{LV}=72.5$ mN$/$m and for our MWCNT samples $F_{adh}=(7.33\pm0.02)$ mN$/$m. Therefore, for a water drop with diameter 1 mm (Figure \ref{fig:Figure2}a), the adhesive force of the MWCNT film in contact with the drop is $F=(7.33\pm0.02)$ $\mu$N. The obtained result is about 25\% lesser than the adhesive force of a single gecko foot-hair (i.e., seta) \cite{Autumn2000},  but 10 times higher than that of \textit{Salvinia} leaf \cite{Hunt2011}. Interestingly, the contact angle value achieved is among the highest reported in literature for not chemically treated \cite{Li2001,Kakade2008a,Wu2010,Wang2010,Clark2012}, functionalized \cite{Yang2010,Kakade2008,Georgakilas2008,Bu2010}, or suitably textured \cite{Lee2010,Lee2011} randomly distributed MWCNT films.\\
\indent Furthermore, Figure \ref{fig:Figure2}b reports the variations of contact angle and droplet radius as functions of the elapsed time from drop cast on the MWCNT films. In such suction experiment we show that although samples are porous, the contact angle trend is constant to demonstrate the stability in time of the super-hydrophobic state of MWCNT coatings. On the other hand, the droplet radius linearly decreases of $\approx15\%$ within 10 min owing to the liquid evaporation and not to the suction process, otherwise the contact angle would also linearly decrease \cite{Gennes2003}. Our results are particularly remarkable, since the water contact angle of MWCNT films has been reported \cite{Huang2005} to decrease linearly with time, from an initial value of $\Theta\approx146^{\circ}$ to $\Theta\approx0^{\circ}$ within 15 min.
\subsection{Salvinia effect in MWCNT films}
\label{sec:SalviniaEffectInMWCNTFilms}
\begin{figure}[ht]
	\centering
		\includegraphics[width=8.6cm,keepaspectratio]{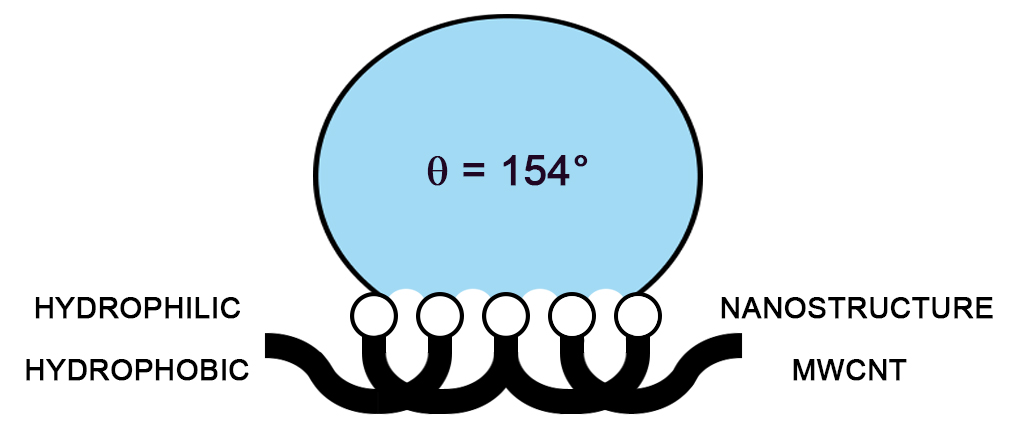}
	\caption{Scheme of the Salvinia effect in our MWCNT random network films. Water droplets are pinned by the attractive interaction due to hydrophilic carbonaceous nanostructures, while they exhibit super-hydrophobic contact angles owing to the repulsive interaction of hydrophobic MWCNTs.}
	\label{fig:Figure3}
\end{figure}

\indent In addition, the obtained high contact angle value may be attributed to hydrophilic ($\approx86^{\circ}$) carbonaceous nanostructures around the tips of the hydrophobic randomly arranged MWCNTs ($\approx92^{\circ}$-$138^{\circ}$ \cite{DeNicola2015a,Wang2010,Clark2012}), constituting a two-fold hierarchical morphology able to stabilize the super-hydrophobicity of the film by the Salvinia effect (Figure \ref{fig:Figure3}). As it occurs in Salvinia, water droplets are pinned by the attractive interaction due to hydrophilic carbonaceous nanostructures, while they exhibit super-hydrophobic contact angles with a large amount of air pockets, owing to the repulsive interaction of hydrophobic MWCNTs. In this way, the film results in a air-retaining super-hydrophobic surface. The super-hydrophobic effect due to the presence of carbonaceous nanostructures on the MWCNT tips has been also reported by Han et al \cite{Han2009}. In their work, vertically aligned MWCNT were processed by plasma immersion ion implantation in order to cap them by hydrophilic amorphous carbon nanoparticles. The authors report a measured water contact angle value of $\approx180^{\circ}$ and zero roll-off angle. However, the difference in contact angle values between our and Han's results could be attributed to the lower density of carbonaceous nanostructures and to the random distribution of the MWCNTs in our samples. Evidently, they observed the lotus effect realizing in that way a waterproof surface, while we recognized the Salvinia effect, thus fabricating an air-retaining super-hydrophobic surface.\\
\indent In order to analyze in more detail the super-hydrophobicity of our samples, we used the Cassie-Baxter equation \cite{Cassie1944} in the hydrophobic regime
\begin{equation}
\cos{\theta^{*}}=\left(1-\phi\right)\cos{\theta}-\phi,\qquad
1=\phi+\phi_{s},
\label{eq:cassie}
\end{equation}
with $\phi_{s}$ the surface solid fraction, $\phi$ the surface air fraction, $\cos{\theta^{*}}$ the apparent contact angle, and $\cos{\theta}$ the Young's contact angle of the surface defined as
\begin{equation}
\cos{\theta}=\frac{\gamma_{SV}-\gamma_{SL}}{\gamma_{LV}}.
\end{equation}
The surface tensions of the solid-vapor, the solid-liquid, and the liquid-vapor interfaces are denoted by $\gamma_{SV}$, $\gamma_{SL}$, and $\gamma_{LV}$, respectively. Moreover, if we consider the experimental contact angle $\theta=97^{\circ}\pm8^{\circ}$ measured for highly pure MWCNT (Nanocyl, NC7000, assay $>$ 90\%, diameter: 5-50 nm) random network films realized as in Ref. \cite{DeNicola2015b}, as the Young's contact angle of the hierarchical MWCNT composite surface with apparent contact angle $\theta^{*}\cong154^{\circ}$, we easily obtain an air fraction $\phi\cong0.88$. This result suggests the formation of a large amount of air pockets. However, we have recently reported \cite{DeNicola2015a} that highly hydrophobic MWCNT random network films with a hierarchical surface morphology, owing to their Young's contact angle close to $90^{\circ}$ are in a metastable Wenzel-Cassie-Baxter state, which is stationary. Indeed, Figure \ref{fig:Figure2}b suggests that air pockets are very stable in time. It is worth noting that the metastability of the MWCNT film is coherent with the Salvinia effect, in which although the liquid droplets are pinned on Salvinia leaves, they can roll off from the surface by a slight vibration.
\section{Conclusions}
\label{sec:Conclusions}
\indent In summary, we have realized super-hydrophobic MWCNT films on AISI 316 stainless steel by CVD without the addition of any external catalysts or pre-treatments, at low-temperature. Furthermore, the investigation at SEM reveals that the MWCNT coatings are carbon nanotube random networks with a two-fold hierarchical morphology owing to the presence of hydrophilic carbonaceous nanostructures on the top of the hydrophobic MWCNTs. The surface hierarchical architecture of the MWCNT films provides a stationary super-hydrophobic state for the coatings because of the Salvinia effect. Such MWCNT films may be used for super-hydrophobic stainless steel realizations, such as drag reduction \cite{Jung2009,Barthlott2010}, anti-corrosion \cite{Shen2005}, anti-fouling \cite{Zhang2005}, and anti-contamination \cite{Koch2009}.
\section*{Acknowledgements}
\noindent We thank R. De Angelis, F. De Matteis, and P. Prosposito (Universit\'a di Roma Tor Vergata, Roma, Italy) for their courtesy of contact angle instrumentation. This project was financial supported by the European Office of Aerospace Research and Development (EOARD) through the Air Force Office of Scientific Research Material Command, USAF, under Grant No. FA9550-14-1-0047.\\

\end{document}